# Energy Spectrum of the Valence Band in HgTe Quantum Wells on the Way from a Two- to Three-Dimensional Topological Insulator

G. M. Minkov[a,b,*], O. E. Rut[a], A. A. Sherstobitov[a,b], S. A. Dvoretsky[c,d], N. N. Mikhailov[c,d], and V. Ya. Aleshkin

[a] *Ural Federal University, Yekaterinburg, 620000 Russia*
[b] *Mikheev Institute of Metal Physics, Ural Branch, Russian Academy of Sciences, Yekaterinburg, 620137 Russia*
[c] *Rzhanov Institute of Semiconductor Physics, Siberian Branch, Russian Academy of Sciences, Novosibirsk, 630090 Russia*
[d] *Novosibirsk State University, Novosibirsk, 630090 Russia*
[e] *Institute for Physics of Microstructures, Russian Academy of Sciences, Afonino, Nizhny Novgorod region, 603087 Russia*
*e-mail: Grigori.minkov@urfu.ru

The magnetic field and temperature dependences and the Hall effect have been measured in order to determine the energy spectrum of the valence band in HgTe quantum wells with the width $d_{QW}$ = 20–200 nm. The comparison of hole densities determined from the period of Shubnikov–de Haas oscillations and the Hall effect shows that states at the top of the valence band are doubly degenerate in the entire $d_{QW}$ range, and the cyclotron mass $m_h$ determined from the temperature dependence of the amplitude of Shubnikov–de Haas oscillation increases monotonically from $0.2m_0$ to $0.3m_0$ ($m_0$ is the mass of the free electron) with increasing hole density $p$ from $2 \times 10^{11}$ to $6 \times 10^{11}$ cm$^{-2}$. The determined dependence has been compared to theoretical dependences $m_h(p, d_{QW})$ calculated within the four-band **k**P model. These calculations predict an approximate stepwise increase in $m_h$ owing to the pairwise merging of side extrema with increasing hole density, which should be observed at $p = (4-4.5) \times 10^{11}$ and $4 \times 10^{10}$ cm$^{-2}$ for $d_{QW}$ = 20 and 200 nm, respectively. The experimental dependences are strongly inconsistent with this prediction. It has been shown that the inclusion of additional factors (electric field in the quantum well, strain) does not remove the contradiction between the experiment and theory. Consequently, it is doubtful that the mentioned **k**P calculations adequately describe the valence band at all $d_{QW}$ values.

## 1. INTRODUCTION

Structure with HgTe quantum wells (QWs) attract great attention for several reasons. First, the QW is formed from a gapless semiconductor, whereas HgCdTe barriers are formed from a semiconductor with normal band ordering.[1] Second, the band structures of the parent materials HgTe and HgCdTe are studied in detail and their parameters are well known. Third, the multiband **k**P method for the calculation of the energy spectrum $E(k)$ in HgTe quantum wells is well developed (see, e.g., [1–4] and references therein). These calculations show that various energy spectra occur depending on the width of the quantum well $d_{QW}$ from the spectrum similar to the spectrum of a narrow-gap semiconductor at $d_{QW}$ < 6.3 nm to a semimetallic spectrum at $d_{QW}$ > ≈15 nm. Fourth, the theory predicts that the HgTe QW with $d_{QW}$ > 6.5 nm is a two-dimensional topological insulator, where one-dimensional edge states are formed in addition to two-dimensional states. The HgTe QW with $d_{QW}$ > 60–80 nm is a three-dimensional topological insulator, where two-dimensional single-spin surface states are formed with the characteristic localization length in the $z$ direction, which is perpendicular to the plane of the QW, much larger than $d_{QW}$. Fifth, the technology of growth of HgCdTe/HgTe/HgCdTe structures is well developed [5, 6].

All these circumstances seem to allow detailed understanding of (transport, optical, etc.) properties of HgCdTe/HgTe/HgCdTe structures.

---

[1] In normal band ordering at the Γ point in II–VI semiconductors, the doubly degenerate $\Gamma_6$ term forms the conduction band, the quadruply degenerate $\Gamma_8$ term constitutes the valence band consisting of band of heavy and light holes, and the doubly degenerate $\Gamma_7$ term forms the spin–orbit split valence band.

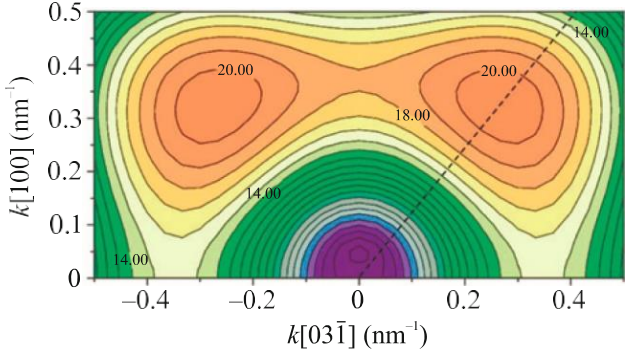

**Fig. 1.** (Color online) Constant-energy contours of the upper split state calculated taking into account the interface inversion asymmetry. Constant-energy contours of the lower split state are similar and located approximately 6 meV below in energy. The calculations were performed for the 8.3-nm-wide quantum well on the (013) substrate at the parameter $g4 = 0.8$. One half of the picture is shown; the second half is a mirror image. The energy is measured from $E(k = 0)$ with a step of 1 meV. Some energies are indicated near the corresponding contours. The dashed line indicates the direction passing through the maximum of $E(k)$.

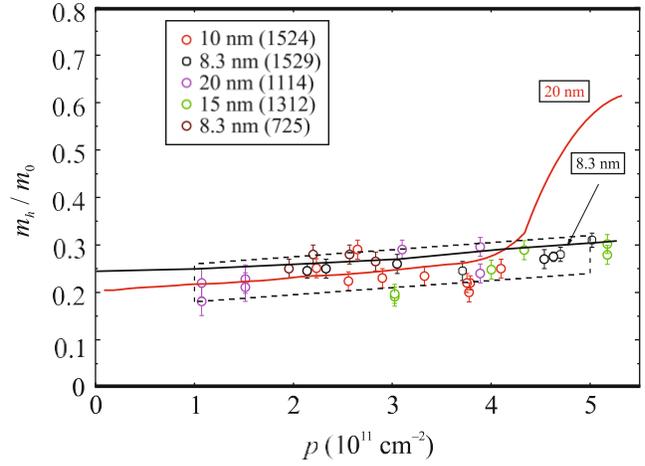

**Fig. 2.** (Color online) Cyclotron mass of holes versus the hole density in structures with $d_{QW}$ = 8.3–20 nm (from [18]). The black and red solid lines are calculated for $d_{QW}$ = 8.3 and 20 nm at $g4L = 0.8$ and $g4R = 1$, respectively. The dashed rectangle marks the $m_h/m_0$ region including all experimental results.

Theoretical calculations predict that the conduction band is quite simple, nearly isotropic and nonparabolic. Its spectrum in structures with $d_{QW}$ = 4–80 nm is thoroughly studied by the optical and magnetotransport methods, as well as by the photoelectric method in a wide photon energy range beginning with the terahertz band [8–12]. It was shown that the energy spectrum is reasonably described in general within the four-band $\mathbf{k}P$ model, and some discrepancies between theory and experiment were discussed in [13].

The energy spectrum of the valence band is much more complex. The theory predicts that the top of the valence band at $d_{QW}$ < 7–7.5 nm is located at $\mathbf{k} = 0$ and has the curvature (mass) close to the curvature of the conduction band. The top of the valence band at $d_{QW}$ > 7–7.5 nm is formed by the four side extrema so that the states at the top of the valence band in symmetric quantum wells have a degree of degeneracy of $K = 8$ (2 owing to "spin" multiplied by 4, which is the number of side extrema). The anisotropy of these states near extrema is small and the cyclotron mass of holes given by the formula $m_h = (h^2/\pi)dS/dE$, where $S$ is the area of the constant-energy cross section at the energy $E$, is $(0.2–0.3)m_0$ at $p < (5–9) \times 10^{11}$ cm$^{-2}$.

The interface inversion asymmetry in HgCdTe/HgTe/HgCdTe structures, which is described by a single parameter $g4$ within the four-band $\mathbf{k}P$ model [14], leads to the "spin" splitting of states at the top of the valence band, so that the degree of degeneracy decreases to 4. Figure 1 presents constant-energy contours of the upper split state calculated taking into account the interface inversion asymmetry in the 8.3-nm-wide quantum well with the parameter $g4 = 0.8$. This figure demonstrates that constant-energy contours of two extrema presented in the figure are merged with a decrease in the energy (i.e., with an increase in the hole density) to 18–18.5 meV, which should lead to the doubling of the cyclotron mass $m_h$ caused by the doubling of $dS/dE$.

The energy spectrum of the valence band is much less studied experimentally [15–17]. It was shown that the extremum of the valence band at $d_{QW}$ = 5–7 nm, when it is located at $\mathbf{k} = 0$, is strongly split owing to the interface inversion asymmetry [15, 17].

At $d_{QW}$ > 7–7.5 nm, when the top of the valence band is formed by four side extrema, the energy spectrum is experimentally studied much less. In [18], we show that the effective mass in the QW with $d_{QW}$ = 8–20 nm at $p = (2–5) \times 10^{11}$ cm$^{-2}$ is close to the theoretical value, but the degree of degeneracy is $K = 2$ rather than 4, as should be the case taking into account the interface inversion asymmetry in the symmetric quantum well. It was shown that the additional asymmetry (e.g., different widths of heterointerfaces or different parameters $g4L$ and $g4R$ characterizing the contribution from the interface inversion asymmetry on the left and right walls, respectively) results in halving of the degree of degeneracy $K$ and, thereby, makes it possible to remove this discrepancy with theory.

Experimental studies of the cyclotron mass of holes in the QW with $d_{QW}$ = 8–20 nm in the hole density range of $p = (2–5) \times 10^{11}$ cm$^{-2}$ showed that the effective mass is close to the theoretical value [17]. The results from [17] are presented in Fig. 2.

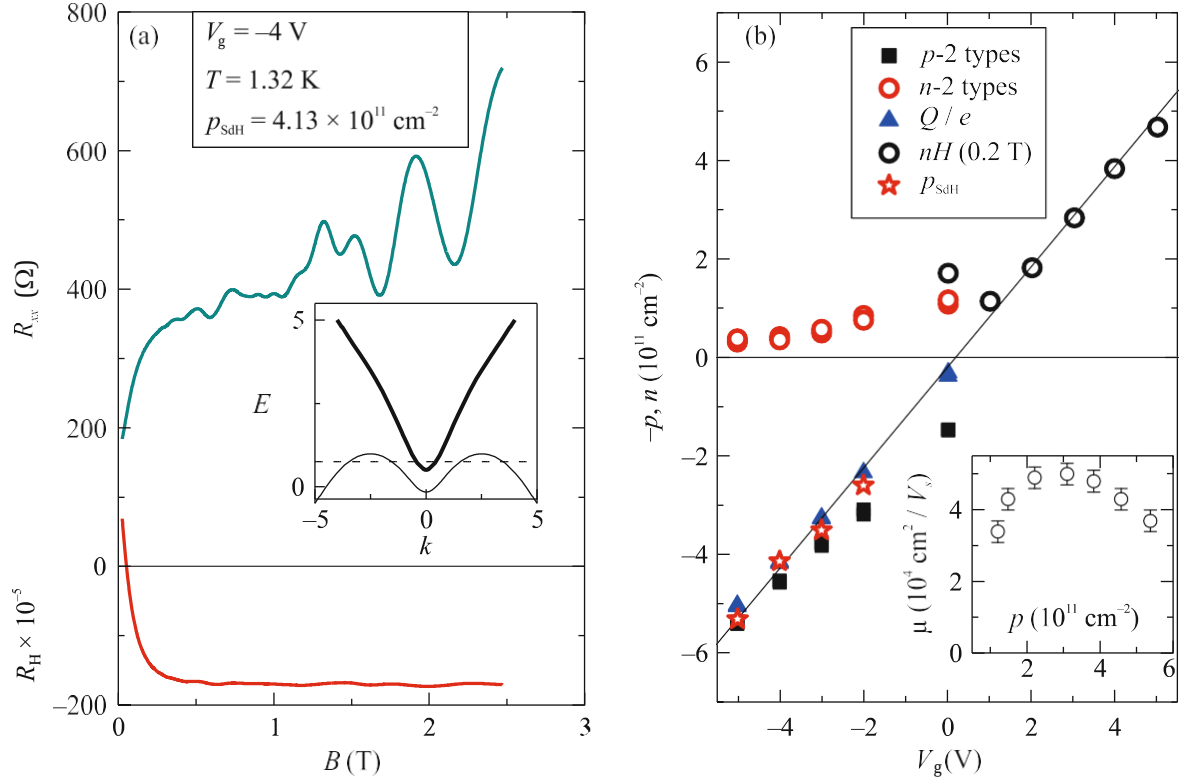

**Fig. 3.** (Color online) (a) Magnetic field dependences of $R_{xx}$ and $R_H$. The inset shows the sketch of $E_c(k)$ and $E_v(k)$ at $d_{QW} >$ 14 nm. (b) Hole, $p$, and electron, $n$, densities and the charge number of the quantum well $Q/e = p - n$ versus the gate voltage $V_g$ determined in the two-carrier conduction model at the gate voltage $V_g < 0$ and the density $p$ determined from the period of Shubnikov–de Haas oscillations under the assumption of double degeneracy of Landau levels. The inset is the mobility of holes versus the hole density.

Any detailed experimental results for $d_{QW} > 20$ nm are absent.

In this work, the effective mass of holes and the degree of degeneracy of the top of the valence band in the QW with $d_{QW} = 20$–200 nm in the hole density range of $(2$–$6) \times 10^{11}$ cm$^{-2}$ are studied experimentally.

## 2. EXPERIMENTAL RESULTS AND DISCUSSION

The studied Hg$_{1-x}$Cd$_x$Te/HgTe/Hg$_{1-x}$Cd$_x$Te ($x = 0.6$–$0.7$) structures with quantum wells of the widths $d_{QW} = 22, 32, 46, 80, 88, 120, 200$ nm were grown by molecular beam epitaxy on the (013) semi-insulating GaAs substrate (in addition, one structure with $d_{QW} = 80$ nm was grown on the (100) substrate). The measurements were carried out with Hall bars with a channel width of 0.5 mm and potential contacts separated by 0.5 mm. An aluminum gate was deposited after the deposition of a gate dielectric (parylene) on the surface of the bars. The dc measurements were performed in the temperature range of 1.3–4 K in magnetic fields up to 5 T.

Experimental results and their processing are identical for all studied structures. We describe them in detail for structure 180824 with $d_{QW} = 32$ nm.

The magnetic field dependences of the longitudinal resistance $R_{xx}$ and the Hall coefficient $R_H$ presented in Fig. 3a show that transport involves at least two types of carriers: electrons, which determine the magnetic field dependences of $R_{xx}$ and $R_H$ in low magnetic fields $B < 0.3$–$0.5$ T, and holes, which determine the magnetic field dependences of $R_{xx}$ and $R_H$ at magnetic fields $B > 0.5$ T (similar dependences are observed at all gate voltages $V_g < 0$). At gate voltages $V_g > 0$, the hole contribution to the conductivity vanishes and $R_{xx}$ and $R_H$ are determined only by conduction electrons. This occurs because the HgTe quantum well with $d_{QW} > 14$–$15$ nm is a semimetal; i.e., the bottom of the conduction band, which is located at the center of the Brillouin zone at $\mathbf{k} = 0$, is below the side extrema in the valence band (see the inset of Fig. 3a). Dependences of the hole and electron densities on $V_g$ are presented in Fig. 3b. The electron density at $V_g < 0$ was determined from the magnetic field dependences of $R_{xx}$ and $R_H$ in the magnetic field range of $B = 0.03$–$0.6$ T in the two-carrier conduction model, whereas the electron density at $V_g > 0$, when $R_H < 0$ and hardly depends on the magnetic field, was determined from the Hall effect at $B = 0.03$ T as $n = (eR_H)^{-1}$, where $e$ is the electron charge.

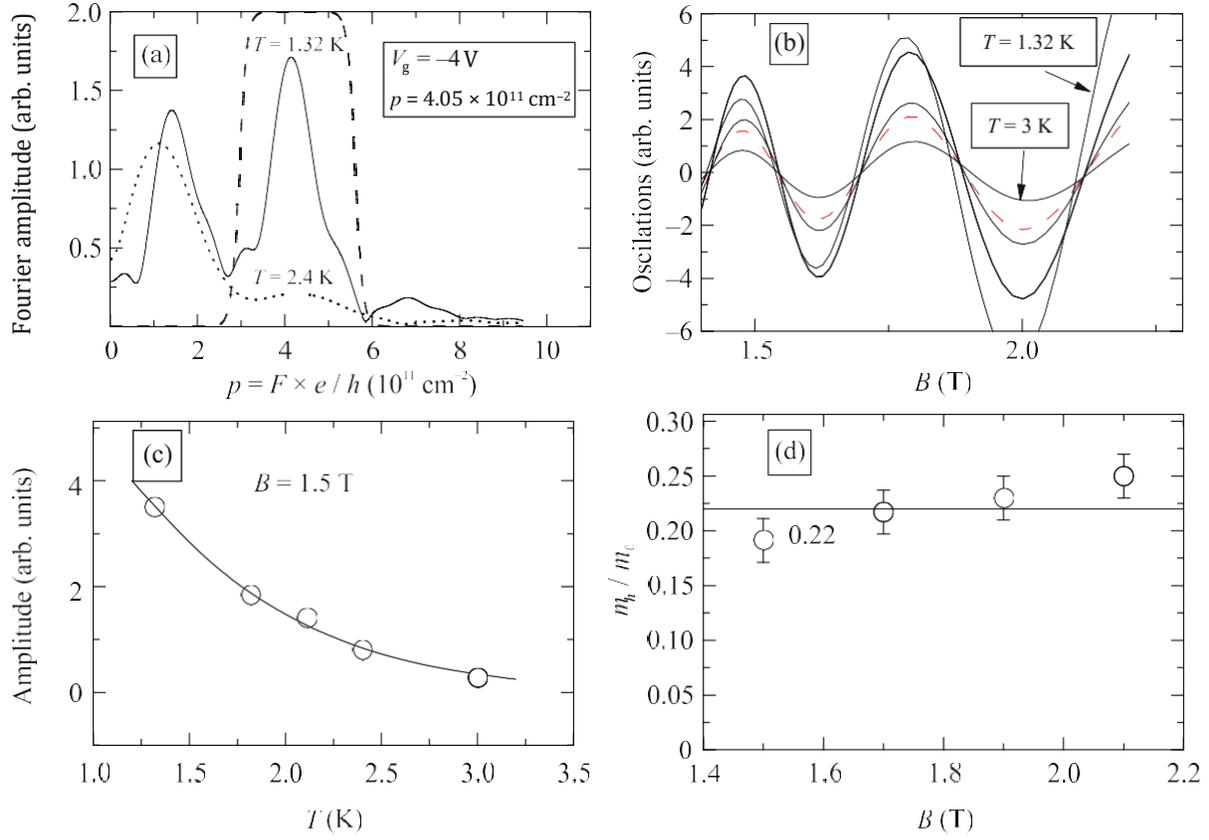

**Fig. 4.** (Color online) (a) Fourier spectrum of oscillations of $R_{xx}$ presented in Fig. 3a; the long-dashed line shows the filter for the separation of contribution from holes to oscillations of $R_{xx}$. (b) Shubnikov–de Haas oscillations of holes found after the filtration of the Fourier spectrum, as shown in panel (a). (c) (Circles) Temperature dependence of the amplitude of Shubnikov–de Haas oscillations in a field of 1.5 T and (line) the Lifshitz–Kosevich formula with $m_h/m_0 = 0.22$. (d) Magnetic field dependence of $m_h/m_0$.

The hole density was determined both from the magnetic field dependences of $R_{xx}$ and $R_H$ in the range of $B = 0.05$–1 T within the two-carrier conduction model and from the frequency $F$ of Shubnikov–de Haas oscillations as $p_{SdH} = (e/h)FK$ (Fig. 4a). Figure 3b demonstrates that $p_{SdH}$ at $K = 2$ coincides within the experimental error with the Hall hole density.[2]

Figure 3b also shows that the charge of the quantum well depends linearly on $V_g$ in the entire $V_g$ range, which indicates the absence of missed conduction channels. This conclusion is confirmed by the fact that the slope of the gate-voltage dependence of $Q/(eV_g)$ coincides within the error with the gate-voltage dependence of $C/S_g$, where $C$ is the capacitance between the two-dimensional gas and the gate and $S_g$ is the area of the gate.

For example, we consider oscillations of $R_{xx}(B)$ in the hole region at $V_g = -4$ V shown in Fig. 3a. The Fourier spectrum of the oscillatory part $\delta R_{xx} = (R_{xx} - R^{mon})/R^{mon}$ of $R_{xx}$, where $R^{mon}$ is the monotonic part of the magnetoresistance, at $V_g = -4$ V is shown in Fig. 4a. The low- and high-frequency components of the spectrum correspond to the contributions from electrons and holes to oscillations of $R_{xx}(B)$, respectively. This immediately follows from the temperature dependence of the amplitudes of these components (Fig. 4a). As the temperature increases from 1.32 to 2.4 K, the amplitude of the low-frequency component decreases only by 20% (this is due to a small effective mass of electrons), whereas the amplitude of the high-frequency component decreases by a factor of about 5.

The first conclusion following from Figs. 3b and 4a is that the degree of degeneracy of Landau levels in the valence band is 2. This follows from the fact that the hole density determined from the period of high-frequency oscillations under the assumption that Landau levels are doubly degenerate coincides within the error with the Hall hole density.

---

[2] This work is focused on the study of the spectrum of the valence band. For this reason, we do not discuss the behavior of $R_{xx}$ and $R_H$ in the electron region. The behavior of this structure in the electron region was analyzed in detail in [18].

To determine the effective mass of holes from the temperature dependence of the amplitude of oscillations, these oscillations were reconstructed by the inverse Fourier transform of the filtered (as shown by the dashed line in Fig. 4a) Fourier spectrum (Fig. 4b). The amplitudes of oscillations in a magnetic field of 1.5 T at several temperatures are presented by circles in Fig. 4c, where the line corresponds to the Lifshitz–Kosevich formula ensuring the best reproduction of the experimental results, which is achieved at $m_h/m_0 = 0.22$. To estimate the error, the ratio $m_h/m_0$ was determined at different magnetic fields. These results are presented in Fig. 4d. Thus, $m_h/m_0 = 0.22 \pm 0.03$ at $p = 4.05 \times 10^{11}$ cm$^{-2}$. Such measurements and their analysis were performed in the entire accessible hole density range; the corresponding results are shown in Fig. 5 together with the calculated dependence $m_h(p)/m_0$.

It is seen that the effective hole mass $m_h/m_0$ at hole densities below $3 \times 10^{11}$ cm$^{-2}$ is in good agreement both with the results for $d_{QW} = 8$–20 nm (Fig. 2) and with the theoretical dependence. However, a sharp increase in $m_h/m_0$ caused by the pairwise merging of side extrema is not observed.

Dependences of $m_h/m_0$ on the hole density calculated with several $d_{QW}$ values in the range of 20–200 nm are presented in Fig. 6 together with experimental results for $m_h/m_0$ obtained only in structures with $d_{QW} = 200$ nm. The experimental $m_h/m_0$ values in structures with $d_{QW} = 22, 32, 46, 60, 88$, and 120 nm lie in the dashed rectangle (experimental values are not presented because they are too numerous and it will be very difficult to understand to which structures different symbols belong).

It is seen that the hole density at which a jump in $m_h/m_0$ occurs because of the pairwise merging of side extrema should decrease strongly with increasing $d_{QW}$ and this jump at $d_{QW} = 200$ nm should be observed at $p = 0.4 \times 10^{11}$ cm$^{-2}$. However, experimental $m_h/m_0$ values at all $d_{QW}$ values are close to each other and increase smoothly from $0.2 \pm 0.03$ at $p = 2 \times 10^{11}$ cm$^{-2}$ to $0.3 \pm 0.03$ at $p = 5 \times 10^{11}$ cm$^{-2}$.

It could be thought that $m_h/m_0$ strongly depends on the orientation of the QW. We tested this assumption for the QW with $d_{QW} = 80$ nm, which was deposited on two substrates with the (013) and (100) orientations. Experimental $m_h/m_0$ values at different hole densities presented in the inset of Fig. 6 show that $m_h(p)/m_0$ is independent of the orientation within the experimental error.

Thus, the degree of degeneracy of states near the top of the valence band is 2 in the entire range $d_{QW} =$

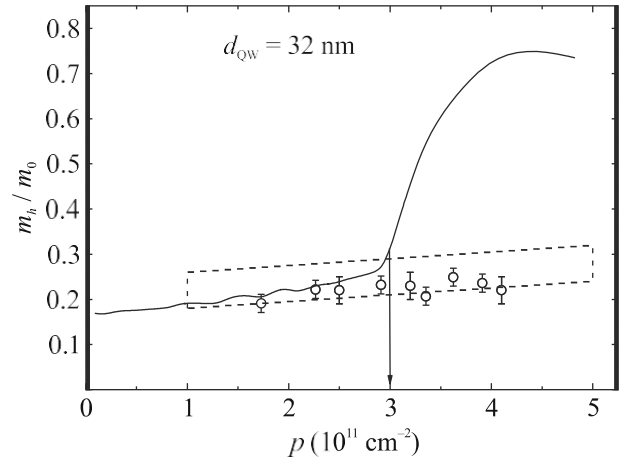

**Fig. 5.** (Circles) Hole mass versus the hole density at $d_{QW} = 32$ nm and (line) the calculated dependence. The arrow indicates the hole density at which side extrema should be pairwise merged, leading to a sharp increase in $m_h/m_0$. The dashed rectangle same as in Fig. 2 indicates the $m_h/m_0$ region including all experimental results at $d_{QW} = 8$–20 nm [18].

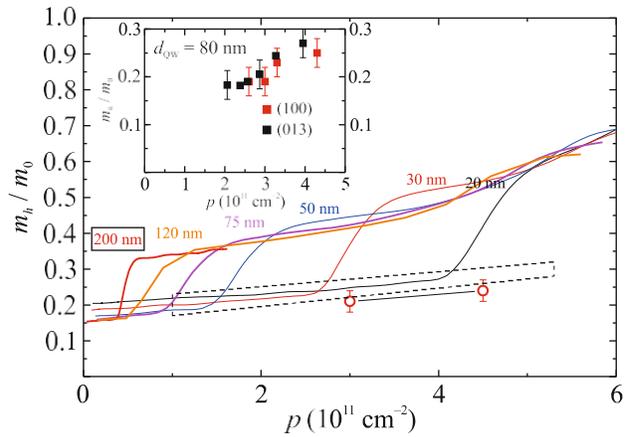

**Fig. 6.** (Color online) (Lines) Calculated density dependences of the hole mass $m_h(p)/m_0$ at the indicated $d_{QW}$ values and (circles) experimental values for the largest quantum well with the width $d_{QW} = 200$ nm. The dashed rectangle same as in Figs. 2 and 5 indicates the $m_h/m_0$ region including all experimental results at $d_{QW} = 8$–20 nm (Fig. 2) and at $d_{QW} = 22, 32, 46, 60, 88$, and 120 nm. The inset shows the experimental $m_h/m_0$ values for two structures with $d_{QW} = 80$ nm on the (013) and (100) substrates.

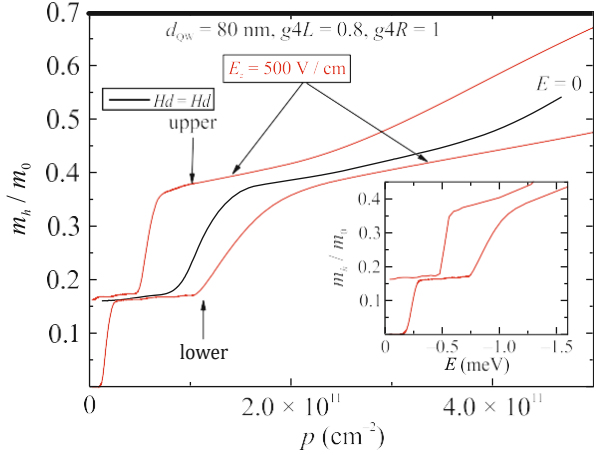

**Fig. 7.** (Color online) Hole mass $m_h/m_0$ versus the hole density in the upper and lower branches of the spectrum split by the electric field $E_z$. The inset shows the hole mass $m_h/m_0$ versus the energy measured from the top of the valence band at different electric field $E_z$.

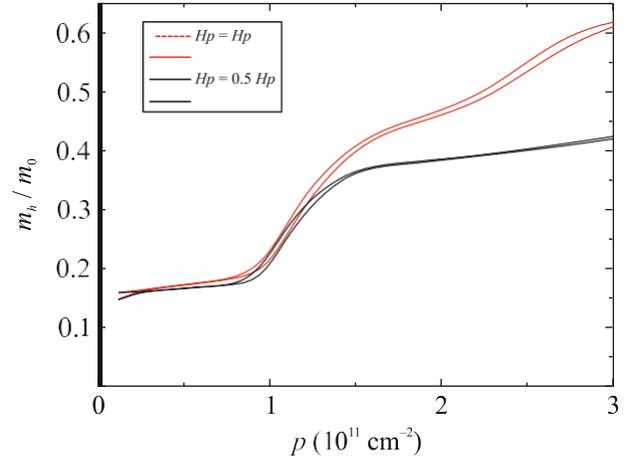

**Fig. 8.** (Color online) Density dependences of the hole mass $m_h(p)/m_0$ at two additions $Hp$ and $0.5Hp$ to the Hamiltonian describing the contribution from deformation.

8–200 nm at hole densities of $(1.5$–$5.5) \times 10^{11}$ cm$^{-2}$. The effective mass of holes $m_h/m_0$ at all $d_{QW}$ values increases monotonically with the hole density from $0.2 \pm 0.03$ to $0.3 \pm 0.03$. This behavior drastically differs from the theoretical spectrum calculated within the four-band $kP$ model, which predicts a stepwise (by a factor of about 2) increase in $m_h/m_0$ at the hole density of $(4$–$4.5) \times 10^{11}$ cm$^{-2}$ in the 20-nm-wide QW and at the hole density of $0.4 \times 10^{11}$ cm$^{-2}$ in the 200-nm-wide QW.

What is a reason for such discrepancy?

**1.** The hole density in the experiment was changed by varying the gate voltage, i.e., in the presence of the electric field $E_z$ in the quantum well, whereas the calculation was performed for the "empty" spectrum. More accurate self-consistent calculations require the $z$ distribution of the charge of holes; i.e., it is necessary to know wavefunctions at energies below the Fermi energy at all $k_x$ and $k_y$ values. This problem seems too difficult and, to estimate the effect of the electric field in the quantum well, we consider $E_z$ = const. The dependences of $m_h/m_0$ on the hole density calculated at $E_z = 5 \times 10^2$ V/cm presented in Fig. 7 show that doubly degenerate states at the top of the valence band are split and the mass jump is shifted toward lower hole densities in one branch and toward higher hole densities in the other branch. In a field of $5 \times 10^2$ V/cm at $p < 2.5 \times 10^{10}$ cm$^{-2}$, only one upper state is filled, so that the degree of degeneracy in this range should be 1 and $m_h/m_0 \approx 0.17$. At $2.5 \times 10^{10} < p < 5 \times 10^{10}$ cm$^{-2}$, both the upper and lower states are filled (marked as upper and lower in Fig. 7), so that the degree of degeneracy in this range should be 2 and $m_h/m_0 \approx 0.17$–$0.18$. At $5 \times 10^{10} < p < 1.2 \times 10^{11}$ cm$^{-2}$, the upper state, where the effective hole mass becomes $m_h/m_0 \approx 0.36$, and the lower state with the effective hole mass $m_h/m_0 \approx 0.18$ are filled. At $p > 2 \times 10^{11}$ cm$^{-2}$, both states with close effective hole masses $m_h/m_0 \approx 0.35$–$0.38$ are filled. Thus, the theory predicts that both the degree of degeneracy and the effective hole mass in the structure with $d_{QW}$ = 80 nm in the presence of the electric field $E_z$ should change with an increase in the hole density. However, at $p > 2 \times 10^{11}$ cm$^{-2}$, as well as in the absence of the field, the degree of degeneracy should be 2 and $m_h/m_0 \approx 0.4$–$0.45$.

This behavior of $m_h/m_0$ is not observed: $m_h/m_0$ remains in the interval of 0.2–0.3 in the entire hole density range. Thus, discrepancy between the theory and experiment cannot be explained by the fact that the calculations presented in Fig. 6 are not self-consistent.

**2.** All calculations were performed under the assumption that deformation in the quantum well remains the same as in narrow wells. However, it can be partially removed in wide wells. To estimate the effect of this factor, we calculated the dependence $m_h(p)/m_0$ at two values of deformation-induced addition $Hp$ to the Hamiltonian: $Hp$ corresponding to the complete deformation (narrow wells) and $0.5Hp$ (Fig. 8).

It is seen that the hole density at which the jump in $m_h/m_0$ should be observed hardly depends on deformation.

Thus, reasons for the drastic discrepancy between experimental and theoretical dependences $m_h(p)/m_0$

remain unclear. Consequently, it is doubtful that the **k**$P$ calculations adequately describe the valence band at all $d_{QW}$ values.

Direct experimental evidence of the existence of four fairly high side extrema is absent. The possibility of achieving agreement between the experiment and theory at $d_{QW} < 20$ nm and $p < 4 \times 10^{11}$ cm$^{-2}$ is not such evidence.

To summarize, the reported study of the energy spectrum of the top of the valence band in HgTe quantum wells with the widths $d_{QW}$ = 8–200 nm has shown that states at the top of the valence band at the hole densities $p < 6 \times 10^{11}$ cm$^{-2}$ are doubly degenerate and the cyclotron mass of holes $m_h$ increases monotonically from $0.2m_0$ to $0.3m_0$ with increasing hole density from $1.5 \times 10^{11}$ to $5.5 \times 10^{11}$ cm$^{-2}$. The dependences $m_h(p)/m_0$ calculated within the four-band **k**$P$ model significantly differ from the corresponding experiment dependences. The estimates of the effects of the electric field $E_z$ in the quantum well and deformation cannot explain discrepancy between experimental and theoretical results. Reasons for this discrepancy remain unclear.

FUNDING

This work was supported by the Ministry of Science and Higher Education of the Russian Federation, project no. 075-15-2020-797 (13.1902.21.0024).

.

REFERENCES

1. L. G. Gerchikov and A. Subashiev, Phys. Status Solidi B **160**, 443 (1990).
2. X. C. Zhang, A. Pfeuffer-Jeschke, K. Ortner, V. Hock, H. Buhmann, C. R. Becker, and G. Landwehr, Phys. Rev. B **63**, 245305 (2001).
3. E. G. Novik, A. Pfeuffer-Jeschke, T. Jungwirth, V. Latussek, C. R. Becker, G. Landwehr, H. Buhmann, and L. W. Molenkamp, Phys. Rev. B **72**, 035321 (2005).
4. Y. Ren, Z. Qiao, and Q. Niu, Rep. Progr. Phys. **79**, 066501 (2016).
https://doi.org/10.1088/0034-4885/79/6/066501
5. C. R. Becker, V. Latussek, G. Landwehr, and L. W. Molenkamp, Phys. Rev. B **68**, 035202 (2003).
6. S. Dvoretsky, N. Mikhailov, Yu. Sidorov, V. Shvets, S. Danilov, B. Wittman, and S. Ganichev, Electron. Mater. **39**, 918 (2010).
7. G. Landwehr, J. Gerschütz, S. Oehling, A. Pfeuffer-Jeschke, V. Latussek, and C. R. Becker, Phys. E (Amsterdam, Neth.) **6**, 713 (2000).
8. X. C. Zhang, A. Pfeuffer-Jeschke, K. Ortner, C. R. Becker, and G. Landwehr, Phys. Rev. B **65**, 045324 (2002).
9. K. Ortner, X. C. Zhang, A. Pfeuffer-Jeschke, C. R. Becker, G. Landwehr, and L. W. Molenkamp, Phys. Rev. B **66**, 075322 (2002).
10. Z. D. Kvon, E. B. Olshanetsky, E. G. Novik, D. A. Kozlov, N. N. Mikhailov, I. O. Parm, and S. A. Dvoretsky, Phys. Rev. B **83**, 193304 (2011).
11. X. C. Zhang, A. Pfeuffer-Jeschke, K. Ortner, V. Hock, H. Buhmann, C. R. Becker, and G. Landwehr, Phys. Rev. B **63**, 245305 (2001).
12. M. S. Zholudev, A. V. Ikonnikov, F. Teppe, M. Orlita, K. V. Maremyanin, K. E. Spirin, V. I. Gavrilenko, W. Knap, S. A. Dvoretskiy, and N. N. Mihailov, Nanoscale Res. Lett. **7**, 534 (2012).
13. G. M. Minkov, V. Ya. Aleshkin, O. E. Rut, A. A. Sherstobitov, A. V. Germanenko, S. A. Dvoretski, and N. N. Mikhailov, Phys. E (Amsterdam, Neth.) **116**, 113742 (2020),
14. E. L. Ivchenko, *Optical Spectroscopy of Semiconductor Nanostructures* (Alpha Science Int., Harrow, UK, 2005), p. 427.
15. Z. D. Kvon, M. L. Savchenko, D. A. Kozlov, E. B. Olshanetsky, A. S. Yaroshevich, and N. N. Mikhailov, JETP Lett. **112**, 161 (2020).
16. A. Yu. Kuntsevich, E. V. Tupikov, S. A. Dvoretsky, N. N. Mikhailov, and M. Reznikov, JETP Lett. **111**, 633 (2020).
17. G. M. Minkov, A. V. Germanenko, O. E. Rut, A. A. Sherstobitov, M. O. Nestoklon, S. A. Dvoretski, and N. N. Mikhailov, Phys. Rev. B **93**, 155304 (2016).
18. G. M. Minkov, V. Ya. Aleshkin, O. E. Rut, A. A. Sherstobitov, A. V. Germanenko, S. A. Dvoretski, and N. N. Mikhailov, Phys. Rev. B **96**, 035310 (2017).